\documentstyle[aasms4,psfig]{article} 
\newcommand{\kms}{{km~s$^{-1}$\,}} 
\newcommand{\zabs}{{$z_{\rm abs}$\,}} 

\def\la{\;
\raise0.3ex\hbox{$<$\kern-0.75em\raise-1.1ex\hbox{$\sim$}}\; }
\def\ga{\;
\raise0.3ex\hbox{$>$\kern-0.75em\raise-1.1ex\hbox{$\sim$}}\; }
\begin{document} 
\title{UVES observations of QSO $0000-2620$: 
Argon and Phosphorus abundances in the dust-free  
damped Ly$\alpha$ system at \zabs = 3.3901   
\footnote{Based on public data from the UVES Commissioning at the
the ESO 8.2m 
KUEYEN telescope 
operated on Paranal Observatory, Chile.}  
} 
\author{Paolo Molaro$^1$, Sergei A. Levshakov$^2$, 
Sandro D'Odorico$^3$,
Piercarlo Bonifacio$^1$, Miriam Centuri\'on$^1$} 
\affil{$^1$ Osservatorio Astronomico di Trieste, 
Via G.B. Tiepolo 11, 34131, Trieste, Italy} 
\affil{$^2$Department of Theoretical Astrophysics, 
Ioffe Physico-Technical Institute, 
194021 St.~Petersburg, Russia}
\affil{$^3$European Southern Observatory, Karl-Schwarzschild-Str. 
2, D-85740 Garching, Germany}
\authoremail{molaro@oat.ts.astro.it} 
\begin{abstract} 
{
The UV resonance transitions of neutral argon 
Ar\,{\sc i}\,$\lambda1066$ \AA\,
and of singly ionized phosphorus 
P\,{\sc ii}\,$\lambda963$ \AA\, 
originated  in the damped Ly$\alpha$ system  (DLA) at
\zabs = 3.3901 towards  QSO 0000--2620 have been
detected by means of the
UVES spectrograph at the 8.2m ESO KUEYEN telescope.   So far,
this is the first measurement of  
Ar\,{\sc i}\, and the second of
P\,{\sc ii}\, ever performed 
in  damped galaxies  and in  high redshift objects.
This DLA is well known for having one 
of the lowest metal abundances and  
 dust content, and the lowest fractional abundance
of molecular hydrogen H$_2$.
The measured Ar abundance is    [Ar/H] = $- 1.91 \pm 0.09$  
which is equal to   the abundances of the other 
$\alpha$-chain elements  (O, S and Si).
The similarity of the Ar
abundance with the other $\alpha$-chain elements implies 
the absence of 
significant photoionization by either UV background  or 
stellar sources along the sightline throughout the damped
Ly$\alpha$ system.

Both  $\log$(Ar/O) and $\log$(Ar/S) ratios are  
found  close to those  measured 
in the extragalactic H\,{\sc ii} regions and in 
blue compact galaxies
where O is more abundant by at least one order of magnitude . 
This strengthens the universality of 
the Ar/O and Ar/S ratios and lends 
support to the existence of a universal IMF.

The abundance of the non-refractory element phosphorus 
  [P/H] = $- 2.31 \pm 0.10$ confirms the
low amount of chemical evolution  in the DLA.   This is 
 the  measurement 
of P in the most metal-poor material and shows 
 a  subsolar   [P/Fe] = --0.27  value.
The measured ratios [P/Si] = $- 0.40 \pm 0.13$
and [P/S] = $- 0.33 \pm 0.13$ provide evidence
for a mild odd-even effect.

Finally, a stringent upper limit 
to the population of the $^3$P$_1$ level in the ground state of
O\,{\sc i} is derived, which provides a lower limit to the
physical dimensions of the  \zabs = 3.3901 system of 
$L > 7$~pc. 
}
\end{abstract}
\keywords{cosmology: observations --- galaxies: 
abundances --- galaxies: evolution
--- quasars: absorption lines} 
\section{Introduction}

 Damped Ly$\alpha$ (DLA) systems are absorption 
systems observed in quasar spectra characterized by large column densities 
($N$(H\,{\sc i})$\ge$ 2$\times 10^{20}$ cm$^{-2}$). 
They span 
the whole epoch of galaxy formation and are generally considered  
  the progenitors 
of the present day galaxies. DLA systems are of particular interest because it
is possible to measure  a variety of chemical elements 
in their interstellar gas-phase
with 
unprecedented precision (Lu et al. 1996; Prochaska \& Wolfe 1999; 
Pettini et al. 1997, 1999 and 2000). However, the observed 
abundance patterns  in the DLAs
 proved difficult  to account for,  due to the possibility
that a significant fraction of some elements contribute  to dust grain
formation  thus depleting   the  gas phase abundances 
(Kulkarni et al. 1997; Vladilo 1998).
At face value the  observed Si, S, Cr, Al, Ni abundances relative to Fe
are similar to those observed
in the old stellar population of the Galaxy, with the notable 
exception of  Zn which is overabundant compared  to iron. 
However, if the [Zn/Fe] {\it anomaly} is interpreted as a
Fe depletion  in the interstellar medium,  the observed abundance pattern
would be similar to   those observed in the warm phase of the interstellar medium
with solar chemical composition. Zinc has not been considered 
a totally 
reliable element by 
Prochaska et al. (2000), because of its uncertain nucleosynthesis. 
A  sample of chemical elements larger than those presently available,
and in particular of elements with little 
affinity with dust grains,
is  required  to better address the
kind of  chemical pattern present in  DLAs.

In this paper we report on the  detection of UV lines of 
the non-depleted species -- 
neutral argon Ar\,{\sc i} and singly-ionized phosphorus
P\,{\sc ii} --
in a damped Ly$\alpha$ absorption (DLA) system at 
\zabs = 3.3901 towards the quasar Q~0000--2620.
  So far,  phosphorus has been measured by Outram et al. (1999)
in the DLA at \zabs=2.625 towards GB 1759+7539 and 
 an upper limit on  phosphorus 
abundance 
has been set for the DLA at \zabs = 2.309 towards
PHL~957 (Molaro et al. 1998).
A possible identification of the Ar\,{\sc i} lines in 
the \zabs = 2.8112 damped Ly$\alpha$ system
towards PKS~0528--250 was reported by 
Srianand \& Petitjean (1998),
however, in their spectrum the Ar\,{\sc i} lines were
considerably contaminated by hydrogen 
absorption which yields a rather uncertain Ar\,{\sc i}
column density. 

The Ar we derive here is the first one 
ever obtained in any DLA and
one of the few values measured by means of
unsaturated lines unlike with the interstellar 
studies in the Galaxy (e.g. Sofia \& Jenkins 1998).
Being  a noble gas with very low condensation temperature,
$T_{\rm c} = 30$~K (Spitzer \& Jenkins 1975),
neutral argon atoms do not 
form chemical bonds with solid compounds, and
the depletion of phosphorus is found to be very small
in the ISM where variations in refractory elements
do occur (e.g. Dufton et al. 1986).
The dust-to-gas ratio in the DLA at z$_{abs}$=3.3901 towards QSO 0000-2620
was found to be 
less than 0.2\% than  that in the
Milky Way (Levshakov et al. 2000b), and thus the measured [Ar/H] and [P/H] 
ratios\footnote{[X/H] $\,\equiv\, \log\,
[N({\rm X})/N({\rm H})] - \log\,[N({\rm X})/N({\rm H})]_\odot$}  
at \zabs = 3.3901 should not be significantly affected
by  dust depletion,
thus  providing a unique probe of
pure nucleosynthetic yields at high redshift.

The study of phosphorus may eventually allow to clarify
the dust formation processes in DLA galaxies. 
Since P and Fe have nearly the same condensation temperature
$T_{\rm c} \simeq 1300$~K, these species should deplete
together in equal proportions if dust is produced in dense
and hot shells of supernova remnants where thermodynamic
calculations of condensation are appropriate. A possible
disparity between   P and Fe gas-phase abundances would favour
a different  model where the dust grains grow in the
interstellar medium (see Jura \& York 1978, and references therein).
 
We discuss here the implications of the Ar and P abundances 
in connection with
the chemical evolution of such damped galaxies.
We underline that 
for research  on the element production history, 
Ar is a quite unique element due to its ability of  tracing
how strongly the apparent abundances of other 
species from H\,{\sc i} regions
could be influenced by ionization.
 
The paper is organized as follows: in \S~2 we briefly describe
observations and analyze  Ar\,{\sc i} and P\,{\sc ii} profiles.
The results on Ar and P abundances are discussed in \S~3
(sections 3.1 and 3.2, respectively).
In Section~3.3, the estimations of the mean gas density and the linear 
size of the
\zabs = 3.3901 system are also given. Conclusions are   summarized 
  in \S~4.

\section{Observations, data reduction and column densities} 

The DLA at \zabs = 3.3901 towards QSO~0000--2620   
shows high neutral hydrogen column density
$N({\rm H\,{\sc i}}) = 10^{21.41 \pm 0.08}$ cm$^{-2}$\, 
(Lu et al. 1996).
This system has been studied extensively since 
it is one of the highest redshift DLAs
seen in the light of a relatively bright QSO 
(Levshakov et al. 1992; Molaro et al. 1996; 
Prochaska \& Wolfe 1999; 
Molaro et al. 2000, hereafter Paper~I;
and Levshakov et al. 2000b, hereafter Paper~II).
 
The spectra of QSO~0000--2620  were  
obtained during the first commissioning of  
the Ultraviolet-Visual Echelle Spectrograph 
(UVES, see Dekker et al. 2000) 
at the Nasmyth focus 
of the ESO 8.2m KUEYEN telescope at Paranal, Chile, 
in October 1999, and have been  released for public use.  
Details of  observations and of  data reduction are given 
in Paper~I.
We recall here that the resolving power is 
${\rm R} = \lambda$/$\Delta\lambda_{\rm instr} \simeq 49000$, 
which corresponds to  
a velocity  resolution of $\simeq$ 6 \kms\, (FWHM).
 
Singly ionized phosphorus, which is 
the dominant state in H\,{\sc i} gas,
has three strong UV lines with $\lambda =  961.041$ \AA,\,
963.801 \AA,\, and 1152.818 \AA\, (the corresponding
$f$-values are equal to 0.3489, 1.458, and 0.2361, according to
Morton 1991).
We have detected with confidence
only the P\,{\sc ii}\,$\lambda963$ \AA\, line which is 
shown in Figure~1   together with the Ni\,{\sc ii}\,$\lambda1709$ 
\AA\, and Fe\,{\sc ii}\,$\lambda1112$ \AA\, lines.
The P\,{\sc ii}\,$\lambda961$ \AA\, is blended and the same 
occurs for  P\,{\sc ii}\,$\lambda1152$ \AA\,
as seen from the EMMI spectrum of Savaglio et al. (1997).
 
Neutral argon has two strong resonance lines in the ultraviolet region
at $\lambda = 1048.220$ \AA\, and 1066.660 \AA. 
Their oscillator strengths  are $f_{1048} = 0.257 \pm 0.013$ and 
$f_{1066} = 0.064 \pm 0.003$ respectively from 
Federman et al. (1992).  The whole region of
the QSO~0000--2620 spectrum  containing  the two Ar\,{\sc i} lines  
is shown in Figure 2 to illustrate  both the crowding of the Ly$\alpha$ forest
and the few windows which were used for the continuum placing.
The Ar lines detected  
at \zabs = 3.3901 are zoomed in Figure~3. 
 
The data used for Ar result from the sum of only 2 spectra instead of the
whole 3  since one of them turned out to be  badly contaminated by a cosmic ray
event.
The S/N ratio for the Ar\,{\sc i}\, $\lambda1066$ \AA\, line therefore is 40.
As can be seen from the figure, the Ar\,{\sc i}\,$\lambda1066$ \AA\,
line is relatively clean, but $\lambda1048$ \AA\ is 
partly blended with additional absorption. 
The Ar\,{\sc i} line strengths
are such that in the Milky Way they tend to be saturated even
for tenuous lines of sight,
but in our DLA system saturation is not an issue  
because of the low metallicity $\bar{Z} \simeq 10^{-2}Z_\odot$
(Paper~I).

The measurements of metal absorption lines falling in the Ly$\alpha$ forest
have to cope with the possible presence  of hydrogen contamination
and with a local continuum, drawn by using  windows
in the Ly$\alpha$ forest which could be located quite far from 
the feature under analysis.

The P\,{\sc ii}\,$\lambda 963$ \AA\, line (Figure~1) shows a rather  
symmetric profile with its blue side reaching the continuum, 
while there could  be a
small contamination in the red wing. The Ar line looks slightly more
 contaminated
with absorption  on both sides. The extra  absorption around the Ar\,{\sc i}\,
line is a 
continuum-like absorption
without signature of discrete
features, which makes it hard  to model with single H\,{\sc i} clouds.
 
Our previous analysis of both metal and molecular hydrogen lines
in this DLA has shown that
the line broadening
in the \zabs = 3.3901 system is mainly caused by 
macroscopic large-scale, rather than thermal, motions. 
All absorption lines from H$_2$ to
Zn\,{\sc ii} show  symmetric profiles with the same 
Doppler parameter of $b \simeq$ 10 \kms\,  within the uncertainty intervals
(Papers I \& II).
The measured broadening parameter, defined 
as $b = \sqrt{2}\,\sigma_{\rm turb}$ where
$\sigma_{\rm turb}$ is the one-dimensional 
gaussian velocity dispersion of ions along the line of sight,
gives a turbulent velocity dispersion 
of about 7 \kms, a  typical value for the Milky Way disk. 
We can therefore assume that the intrinsic profile
of the Ar\,{\sc i}\,$\lambda1066$ \AA\, and P\,{\sc ii}\,$\lambda 963$ \AA\, lines
is the same for  all neutral atoms and low ionized
species from the H\,{\sc i} region since they
trace the same
volume elements. 

To calculate the confidence range for the Doppler parameter
we make use  of all   metal lines associated with 
the DLA redwards the Ly$\alpha$ emission and minimize
$\chi^2$ by varying all the other  parameters.
The calculated $\chi^2_{\rm min}$ values
as a function of $b$ with the $1\sigma$ and $2\sigma$ confidence levels are shown
in Figure~4. At $\chi^2_{\rm min}$ = 1.148 the most
likely value for $b$ is 10.00$^{+0.29}_{-0.25}$ km s$^{-1}$.
The confidence ranges are computed following the 
procedure described in Press et al. (1992), here we fix the value of 
$b$ and minimize
$\chi^2$ by varying all the other  parameters.

The P\,{\sc ii}\, $\lambda963$ \AA\, and Ar\,{\sc i}\, $\lambda1066$ \AA\, lines are fitted with
a theoretical one-component Voigt profile to the observed
intensities together with the full set of 6 metal lines, for 5 ionic species,
which are located redwards the Ly$\alpha$ emission,
namely the lines  Zn\,{\sc ii}\, $\lambda 2026.1360$ \AA\,,
Cr\,{\sc ii}\, $\lambda2062.2339$ \AA\,, $\lambda2056.2539$ \AA\,, 
Fe\,{\sc ii}\, $\lambda1611.2004$ \AA\,,  Si\,{\sc ii}\, $\lambda1808.0126$
\AA\, and Ni\,{\sc ii}\, $\lambda1709.600$ \AA\,, 
assuming that $b$ is the same for all lines and leaving all
the elemental column densities as well as the local continuum 
$(\Delta C/C)$ around the 
P\,{\sc ii}\,$\lambda963$ \AA\, and Ar\,{\sc i}\,$\lambda1066$ \AA\, lines
free to vary. We have therefore a total of 7 free parameters and 
111 and 105 points for P\,{\sc ii}\, and Ar\,{\sc i}, respectively,
thus giving for the former $\nu = 104$ 
degrees of freedom, and 98 for the latter.
All the lines have been previously adjusted in wavelength
by using the 
value of \zabs = 3.390127 which corresponds to the
redshift of the isolated Ni\,{\sc ii}\,$\lambda 1709$ \AA\,
absorption line (Paper~I). 
The calculated 
theoretical profiles were convolved with a gaussian
instrumental profile with   6 \kms width.
 
The best fit for P\,{\sc ii}\, gives
$N({\rm P}\,{\sc ii}) = 4.29(\pm0.39)\times 10^{12}$ cm$^{-2}$, $b$ = 9.96  \kms, and 
$\Delta C/C = -0.0778$ for a $\chi^2_{\rm min}$ = 1.10.
The result   
is shown in Figure~1 by the grey solid curve, whereas points
with error bars give the normalized intensities,
corresponding to a signal-to-noise ratio S/N $\simeq 15$ per pixel.
Note that the intensities in the bottom panel of Figure~1 are re-normalized in 
concordance with the 
local $\Delta C/C$.
The dotted curve  shows the un-convolved
theoretical P\,{\sc ii} profile in the range
$|\Delta v| \leq 40$ \kms .
The column densities of Zn, Cr, Si, Ni and Fe are
consistent within 0.02 dex with those found in Paper I.
The  errors in the P\,{\sc ii}\, column density are
of 0.04 dex as can be deduced from Figure~5 (panel a), where the 
$\chi^2_{\rm min}$  as a function of the $N$({\rm P}\,{\sc ii}) 
column density is  shown.  
The variations for $\Delta C/C$, 
as well as $b$ as a function of $N$(P\,{\sc ii}\,)  are also shown
in panels (b) and (c), respectively.
The grey area in panel (b) marks the range of possible continuum
deviations which correspond to the $1\sigma$ confidence interval
shown in panel (a).

The analysis of  Ar relies entirely on  Ar\,{\sc i}\,$\lambda1066$ \AA\, 
while  Ar\,{\sc i}\,$\lambda1048$ \AA\, of which only a portion is  
unblended, 
is checked for consistency with the line model but it is not used 
in the fit. 
The corresponding signal-to-noise ratio in this 
spectral region is approximately equal 
to 40.
Figure~3 shows the best fit for  the one-component profile 
(grey continuous curve)
and the intrinsic un-convolved profile for only  the
Ar\,{\sc i}\,$\lambda1066$ \AA\, line  (dotted curve).
This figure demonstrates that the best fit obtained for
the  Ar\,{\sc i}\,$\lambda1066$ \AA\, line 
provides a consistent 
fit of the red wing of 
the stronger Ar\,{\sc i}\,$\lambda1048$ \AA\,  line.
Note that a decrease of 0.1 dex in the Ar\,{\sc i}\, column density is 
sufficient to
destroy the perfect match present  in the Ar\,{\sc i}\, $\lambda1048$ \AA\, line.
A $\chi^2_{\rm min}$ = 1.138 is found for 
$N({\rm Ar}\,{\sc i}) = 1.044(\pm 0.069)\times10^{14}$ cm$^{-2}$, 
$b$ = 10.05 \kms and $\Delta C/C = -0.0086$.
Figure~6 shows  the  confidence levels for the Ar\,{\sc i}\, column density, 
the  variations for  $\Delta C/C$, 
and the Doppler broadening.   

To test  the whole procedure
we analyzed the  Fe\,{\sc ii}\,$\lambda1112$ \AA\, line, also 
shown in Figure~1.
This is the only Fe\,{\sc ii}\, line 
in the Ly$\alpha$ forest which is not strongly blended.
We then  compared the results  
with 
those of  the Fe\,{\sc ii}\,$\lambda1611$ \AA, which is
placed  
redwards the QSO Ly$\alpha$ 
emission in a clean  spectral region  and for 
which a reliable iron column density has already 
been derived in Paper~I.
For 
the Fe\,{\sc ii}\,$\lambda1112$ \AA\, line 
we used 
the empirical oscillator strength $(f_{1112} = 0.0062)$ 
derived recently by Howk et al.
(2000) by means of FUSE observations.
A  $\chi^2_{\rm min}$ = 1.0299 is found for 
$N({\rm Fe}\,{\sc ii}) = 7.095(\pm 0.875)\times10^{14}$ cm$^{-2}$, 
$b = 10.07$ \kms and $\Delta C/C = -0.066$.
In this case we used only 6 metal lines from the red part of the quasar
spectrum excluding Fe\,{\sc ii}\,$\lambda1611$ from the 
$\chi^2$-minimization procedure. Thus we had 6 free parameters
\{$N_{{\rm Ni}\,{\sc ii}}, N_{{\rm Si}\,{\sc ii}}, N_{{\rm Zn}\,{\sc ii}},
N_{{\rm Cr}\,{\sc ii}}, b, \Delta C/C$\}, 93 data points, and
$\nu = 87$ degrees of freedom, and calculated $\chi^2_{\rm min}$
as a finction of $N({\rm Fe}\,{\sc ii}\,\lambda1112)$.
The calculated confidence intervals are shown in Figure~7.
The mean iron column density for Fe\,{\sc ii}\,$\lambda1112$ \AA\, is   
0.02  dex 
lower than that derived from  Fe\,{\sc ii}\,$\lambda1611$ \AA\,  
in Paper~I
and is fully consistent within  errors.
This  shows that we can indeed  recover the ion column density accurately
even though the line is falling in the Lyman $\alpha$ forest with
random Ly$\alpha$ interlopers and uncertainties associated with the 
continuum drawing.        
The derived column densities and abundances for 
Ar and P, together with
those obtained for other species in previous studies, 
are reported in Table~1.

The results presented above differ quite slightly 
from those obtained from a more commonly used  analysis in which 
the continuum is kept fixed while both the broadening and the column density
are left free to vary.
In this approach we obtain
for phosphorus  
$N$(P\,{\sc ii}) $= 10^{12.70 \pm 0.04}$ cm$^{-2}$
and $b = 10.3^{+1.4}_{-1.1}$ \kms, while for argon 
$N$({\rm Ar}\,{\sc i}) =
$10^{14.02 \pm 0.02}$ cm$^{-2}$ and the $b = 9.95 \pm 1$ \kms 
when the P\,{\sc ii}\,$\lambda963$ \AA\, and 
Ar\,{\sc i}\,$\lambda1066$ \AA\, lines were analized separately.
The shift of the P\,{\sc ii} line
is $\Delta v = - 0.01$ \kms, and that of the Ar\,{\sc i} line center is
negligible, namely $\Delta v = - 0.66$ \kms, which is less
than one-half of the pixel width.

The fact that both lines have the same centers and the same
$b$-parameters
which are very similar to those of metal lines
observed in the low line density portion of the 
spectrum redwards the Ly$\alpha$ emission line,
shows that the contamination 
of  Ar\,{\sc i} and P\,{\sc ii}  lines observed
in the Ly$\alpha$ forest is not significant in 
the \zabs = 3.3901 system towards QSO 0000--2620.

\section{Discussion}
 
For  both argon and phosphorus
there are no
 determinations
of the intrinsic abundances in either dwarf or giant
stars because argon has optical transitions
from very high excitation levels, and phosphorus shows
too faint lines in stellar spectra. 
P was measured in chemically peculiar stars (Castelli et al. 1997), 
halo horizontal-branch stars by
Bonifacio et al. (1995), and sdB  by  Ohl et al. (2000). 
However, the  elemental  abundances in these stars are likely modified by
atmospheric processes such as diffusion.
A firm identification of the P\,{\sc v} and P\,{\sc iv} has been made by 
Junkkarinen et al. (1997) in the BAL QSO PG 0946+301 and 
a  remarkable overabundance of phosphorus
with P/C $\approx$ 60(P/C)$_{\odot}$ was found.

The lack of such  important observational 
reference complicates
the interpretation of the Ar and P abundances
in the DLA systems.
Nevertheless, we may compare 
the measured [Ar/H] and [P/H] ratios
with theoretical predictions based on integrating 
massive star yields over an initial mass function 
 given, for instance, by Timmes et al. (1995,
hereafter TWW).  
In this section we consider such issue and other
physical properties of the \zabs = 3.3901 system.
 
\subsection{Argon Abundance}

\subsubsection{Local stellar ionization}

Argon is an  extremely  volatile element. 
It has been detected and studied in the Galactic ISM 
by means of Copernicus, IMAPS and recently by FUSE.
The gas-phase (Ar/H) ratio
in the Galactic interstellar medium is found to be 
below its solar value with reduction factors
varying with the lines of sight.
Fitzpatrick (1996) argued that abundances 
of undepleted elements in the ISM
such as C, N, and O are below their solar values, 
but this cannot explain the line of sight variations
and the considerable amount of the observed
Ar deficiency, [Ar/H] $\la -0.5$ dex.
Since Ar can be hardly  depleted into dust grains, 
Sofia \& Jenkins (1998) suggested 
that the apparent Ar deficiency is 
caused by a partial ionization of Ar by nearby
UV stellar radiation.

Ar\,{\sc i} has an extraordinarily 
high photoionization cross section 
for photons with energy higher than the IP = 15.76 eV.
Its photoionization cross section is about ten times higher than
that of H\,{\sc i} over a broad range of energies.
Therefore, for a cloud that is not thick enough to shield its
interior from outside sources of ionizing radiation,
the UV photons may penetrate the cloud and partially ionize 
hydrogen, argon and other elements. In this case the fraction
of Ar\,{\sc i} can be lower than the corresponding fraction
of H\,{\sc i}
since the absorption cross section of hydrogen 
diminishes $\propto \nu^3$ for energies 
$h\nu > 13.6$~eV, an effect which may lead to an apparent
deficiency of argon with respect to hydrogen if one assumes
that both elements are mostly  neutral.

Reversing the argument presented by Sofia \& Jenkins  
on the Ar\,{\sc i} 
deficiency in the local interstellar medium,
the coincidence of the Ar abundance with those of O, Si and S 
(see Table~1)
implies the absence of significant photoionizing flux from stars
along the line of sight throughout the 
\zabs = 3.3901 damped galaxy.
Otherwise,
the presence of stellar photons with $h\nu > 13.6$~eV
would decrease the Ar\,{\sc i} abundance.
The same effect is not observable in O\,{\sc i}, 
although neutral oxygen also has  a photoionization 
cross section larger than that of H\,{\sc i}.
Oxygen ionization fractions are coupled with  those of hydrogen 
via resonant charge exchange reactions. 
Therefore, there is  observational evidence 
 that in this particular damped system 
the observed ionic species are from
their dominant ionization stages in the H\,{\sc i} gas and,
hence, their abundances do not require ionization 
corrections which may be caused by a hard radiation
field in a uniform  and  
fluctuating 
gas density absorbing region, respectively (Viegas 1995, Levshakov et al. 2000a).
We would like to emphasize once more that
our measurement of Ar\,{\sc i} rules out 
the presence of active star formation regions
and of significant partly ionized gas along 
the line of sight at \zabs = 3.3901.

Steidel \& Hamilton (1992)  identified a galaxy at $2.8^{\prime\prime}$ 
from the QSO~0000--2620 sightline,  which was supposed to be responsible for the
\zabs = 3.3901 damped system.
Giavalisco et al. (1994) found no Ly$\alpha$ flux in correspondence 
of this object by means
of narrow-band imaging placing a 3$\sigma$ upper limit of 
$1.2\times10^{-17}$ erg s$^{-1}$ cm$^{-2}$. 
The object identified in Steidel
\& Hamilton (1992) is probably a high redshift galaxy at $z \sim 2.8$
(Steidel 2000), and therefore  the galaxy at $z =3.39$  still remains undetected.
However, the upper limit provided by Giavalisco et al. (1994)
in the Ly$\alpha$  emission refers to the sky area around the QSO 
and remains valid regardless of the nature of the DLA candidate.
The general lack of Ly$\alpha$ emission from DLAs 
has been commonly explained 
by the effect of dust and resonant scattering which can 
attenuate Ly$\alpha$ photons (Charlot \& Fall 1991, 1993). 
However, the 
extremely low dust content
in our system with a dust-to-gas ratio of less than  10$^{-3}$ of 
the   Milky Way value,
coupled with no detectable effects on the Ar\,{\sc i} abundance,  suggests that 
most likely the
DLA in question is not a site of intense star formation.

\subsubsection{A universal IMF}
 
Observations of emission-lines in ionized  H\,{\sc ii} regions of spiral 
and irregular galaxies provide
abundances for O, S and Ar among the others. 
None of these elements is expected to be
heavily depleted into the solid phase as  dust
(Savage \& Sembach 1996). 
It has already been pointed out that the abundance ratios 
[S/O] and [Ar/O] appear 
to be universally constant and independent of metallicity  
over about 2 decades of oxygen abundances (e.g. Henry \& Worthey
1999). Since O production depends on the progenitor mass while Ar and S do not,
the implication of this constant ratio is 
that either the initial mass 
function is universally 
constant or that the observed elemental ratios 
are not sensitive to IMF variations. 
 
Sulfur and argon originate from 
explosive oxygen burning in Type~II supernova events 
through the chain $^{16}$O + $^{16}$O $\rightarrow$
$^{28}$Si + $^4$He, then
$^{28}$Si + $^4$He $\rightarrow$ $^{32}$S, and
$^{32}$S + $^4$He $\rightarrow$ $^{36}$Ar. 
However, substantial amounts of S and Ar may be manufactured 
in Type~Ia supernova events (Nomoto et al. 1997). 
Considering the values of Izotov \& Thuan (1999) for the 
extragalactic H\,{\sc ii} metal-poor regions,
we have $\log ({\rm S/O}) = -1.55 \pm 0.06$ (rms) 
and $\log ({\rm Ar/O}) = -2.25 \pm 0.09$ (rms).
Similar results, $\log ({\rm S/O}) = -1.48 \pm 0.06$ (rms) and
$\log ({\rm Ar/O}) = -2.32 \pm 0.05$ (rms), for the same class of objects 
have been
recently obtained by Kniazev et al. (2000).
 
These observed ratios are comparable with the 
theoretical predictions of 
Woosley \& Weaver (1995), 
Nomoto et al. (1997), and  Samland (1998) 
in which  log\,(S/O) is $\approx$ --1.75, --1.9, --1.6, and  log\,(Ar/O) $\approx$ 
 --2.45, --2.7, and  --2.4,
respectively. 
The theoretical predictions fall 
somewhat below the observed range,
and either an overproduction of O in the models or 
a significant production of S and Ar in Type~Ia
supernovae are advocated.
 
Our results listed in Table~1 show that 
the Ar abundance follows closely the other $\alpha$-chain
element abundances in the \zabs = 3.3901 system.
The absolute  oxygen abundance in this DLA is 
$\log\,({\rm O/H}) + 12 \approx 7$, i.e. one order
of magnitude below the value obtained in the most metal-poor galaxies
where these elements were measured.
The measured log\,(S/O) and log\,(Ar/O) ratios are of 
$-1.72 \pm 0.15$ and $-2.40 \pm 0.15$,  respectively, i.e. 
very close to the 
observed ratio in  blue compact and spiral galaxies.
This strongly favors  
the universality of (S/O)
and (Ar/O) ratios. 
The implication pointed out by Henry \& Worthey (1999)
that either the IMF is universally constant or  
the stellar mass range which 
produces these elements is 
narrow enough to
make the ratios  insensitive to IMF variations
is reinforced and further 
extended to much more deficient material
detected at \zabs = 3.3901.
 
\subsubsection{$\alpha$-capture versus Fe-peak element ratios}

Most of $\alpha$ particle nuclei are synthesized primarily by 
Type~II supernovae during
the early stages of the galaxy chemical evolution,
while the Fe-group elements are
primarily produced by Type~Ia supernovae and build up 
over  longer time scales.
As a result, in the early phases of chemical enrichment 
the $\alpha$-chain elements  
are expected to be relatively more abundant than the 
iron-peak elements.
 
Argon is a typical product of Type~II SNe, and 
like zinc it is not depleted in the interstellar medium. 
This makes the [Ar/Zn] ratio  
a new and  interesting dust-free diagnostic tool for   
verifying the presence of the $\alpha$-chain element enhancement. 
By combining the present abundances with 
those previously determined (see Table~1)
we have  disposal of  a set of 4 indicators for the 
   $\alpha$-chain element abundances
such as O, Si, S, Ar, and
4 indicators for the abundances of  iron-peak elements
such as Cr, Fe, Ni, and Zn. 
This is a rather unique circumstance among the DLA systems
studied so far.
 
The measured [O,Si,S,Ar/Cr,Fe,Ni,Zn] ratios reported  
 in Table~2 for quantitative comparison
provide 
$\langle[\alpha$-chain/iron-peak]$\rangle = 0.18 \pm 0.03$ 
dex (the weighted mean and dispersion)  which  becomes 
$0.12 \pm 0.02$ when 
the slightly deviating Ni is not considered.
  In the case that   Ni  resulted  a better  Fe-peak indicator then Fe, Cr and 
Zn, we would indeed find
that [O,Si,S,Ar/Ni] are  +0.41, +0.36, +0.29 and  +0.36,
respectively. However, Prochaska \& Wolfe (1999) measured  Ni\,{\sc ii}\, 
$\lambda1751.910$ \AA\,
and found [Ni/H] = $-2.335 \pm 0.085$, which becomes 
[Ni/H] = $-1.952 \pm 0.085$
when the $\log gf$ values are taken from  Fedchak \& Lawler (1999) as we did 
for the Ni\,{\sc ii}\, $\lambda1709.600$ \AA\, (Paper I). The slightly 
different abundance obtained 
from the two Ni\,{\sc ii}\, lines
prevents Ni from being considered a safe proxy to the iron-peak abundances.

An  [alpha/iron-peak] ratio of $\approx$ 0.1  is significantly
lower than similar  quantity in 
Galactic stars with  comparable metallicities, 
[O,Si,S/Fe]~$\simeq 0.3-0.6$ dex (McWilliam 1997, and
references therein). The difference would be even more striking if the 
high O values of 
[O/Fe] = 0.7 found in halo stars at [Fe/H] = --2 by Israelian et al. (1998) 
and 
Boesgaard et al. (1999) were  confirmed.
Implications of low $\alpha$-chain/iron-peak 
element ratios are discussed in Paper~I,  
and the reader is referred to it for additional  details.
Here we notice that the lack of significant [Ar/Zn] 
enhancement corroborates our previous 
results on this system as well as  
the assumption that, on the basis of their relative element
abundances, at least some of the DLA galaxies seem to undergo a 
chemical evolution which
differs from that of the Milky Way (Centuri\'on et al. 2000).  
An  analysis of the zinc abundances in the thick disk stars  
by Prochaska et al. (2000)  showed
that zinc could be
mildly overabundant with respect to iron with [Zn/Fe] = $0.09 \pm 0.023$, 
a slight difference from the thin disk and halo behaviour found by 
Sneden et al. (1991). 
However, in our DLA  
[Zn/Fe] = $-0.03 \pm 0.06$ and [Zn/Cr] = $-0.08 \pm 0.06$,  which   
suggests  that in the DLAs zinc tracks   
Fe and Cr closely.
 
\subsection{Phosphorus Abundance}

Phosphorus is likely  a non-depleted element.
Dufton et al. (1986)  and Jenkins et al. (1986) found that
phosphorus is essentially undepleted along sightlines 
sampling  low density
neutral gas and is depleted by approximately 0.5 dex 
in cold clouds.
However, P is found  strongly depleted 
in dense star forming molecular clouds
(Turner et al. 1990).
From FUSE observations, Friedman et al. (2000) 
obtained [Fe/P] $\approx -0.6$  and --0.7
for 
the Milky Way and LMC diffuse interstellar clouds, respectively, 
and argued for a significant incorporation of 
iron into dust grains in both galaxies.
The phosphorus abundance provided here is one of the few ones
available for this element in any astronomical site and 
the only one in  little processed
material.

In our DLA the abundances of the refractory elements
Fe and Cr are  close
to that of the non-refractory element Zn, showing the absence 
of significant dust concentration 
along the line of sight (Paper~I).
Such a circumstance is extremely rare among DLAs since
even in the cases 
with low element depletion considered by Pettini et al. (2000),
a difference between Zn and 
Fe, Cr and Ni has been always  observed,  with the exception of 
the damped system towards  QSO 0454+039 where [Zn/Cr] $\approx$ 
0 within the errors.
Moreover, the extremely low fraction of molecular hydrogen  
$f({\rm H}_2) \equiv 2\,N({\rm H}_2)/N({\rm H})_{\rm total}
\simeq 4\times10^{-8}$ in our DLA (Paper~II) rules out
a significant depletion of P in 
a variety  of molecular compounds.
Taking into account that the absorbing gas is mostly neutral,
as discussed above, we  conclude that
the observed gas-phase P abundance reflects
the total amount of P.

The low abundance of P is close to those of other elements 
(see Table~1) and supports  independently
the early suggestion  
that the \zabs = 3.3901 galaxy has indeed a low metallicity.
In the context of the other published  data  this result
suggests  a possible mild decrease  
in the damped abundances
at high redshift, which, however, 
is not evident when the column density-weighted
mean is considered  (Prochaska \& Wolfe 2000; Vladilo et al. 2000).

Theoretically the processes that  produce the odd-Z element 
phosphorus in late stages of 
stellar evolution are not 
identified clearly. As Margaret Burbidge (1999) pointed out
the P synthesis
was virtually ignored by Burbidge et al. (1957). 
The only stable isotope is $^{31}$P, which  is     
strongly underproduced in the explosive O burning. P is
likely produced in the carbon and neon shell burning
by neutron capture similarly to the parent nuclei $^{29}$Si and $^{30}$Si.
No significant P production is found during the 
explosion phases (Woosley \& Weaver 1995).
 
Phosphorus abundance, like that of the other odd-Z elements Na and Al,
should be proportional to  $\eta$, 
the total neutron excess in a given amount of matter\footnote{
the neutron excess is defined as
$\eta = (n_{\rm n} - n_{\rm p})/(n_{\rm n} + n_{\rm p})$, where
$n_{\rm n}$ represents the total number of neutrons per gram, both
free and bound in nuclei, and $n_{\rm p}$ the corresponding number for
protons (Arnett 1971).},
which is $\eta \approx 1.5\times10^{-3}(\zeta/\zeta_{\sun})$, 
where $\zeta_{\sun} \approx 0.02$ denotes the fraction by mass of matter
composed of nuclei with atomic number Z~$\ge 6$
(Arnett 1971, 1996).
As the metallicities decrease,  the available nuclei which provide the 
neutron excess decrease too. 
However, in the  primordial 
stellar yields Arnett (1996) found that the 
odd-Z nuclei are underabundant with reference  to the
even-Z neighbours, but the effect is much smaller  than that predicted by simple
arguments of linear increase of neutron excess with metallicity.
Theoretical models by
Woosley \& Weaver (1995) lead  also to substantial 
P production with P yields showing a metallicity dependence 
which is more pronounced at very low metallicities. 

In our DLA the measured ratios   [P/Si] = $- 0.40 \pm 0.13$
and [P/S] = $- 0.33 \pm 0.13$ provide evidence
for a mild odd-even effect, which, however, is  less than what expected on the
basis of the Woosley \& Weaver (1995) and Limongi et al. (2000) yields.
  A similar value for [P/S]=$-0.34 \pm 0.07$  has been derived by
 Outram et al. (1999)
in the DLA at \zabs=2.625 towards GB 1759+7539, which with [Fe/H]=--1.31
is however
somewhat more metal rich than the DLA studied here.

In stellar spectroscopy the element abundances
are traditionally discussed as related  to iron, [X/Fe], and
as a function of the iron-to-hydrogen ratio, [Fe/H].
Following this way, in Figure~8 we compare  our results
with theoretical predictions since 
in our case there is no significant
dust depletion in the \zabs = 3.3901 absorber
and, hence, [Fe/H] can be used as a reference ratio. 
Figure~8 shows the result
of such comparison between the elements listed in Table~1
(dots with 3 $\sigma$ error bars)
and the theoretical prediction (shaded area). 
The width of the shaded area just above a marked element
shows the uncertainty range ($\pm 0.2$ dex) 
for the relative abundance of this element [X/Fe] at
fixed metallicity [Fe/H] = -- 2.0.
The uncertainty of  theoretical predictions is
caused by unknown model parameters as described in TWW.
The upper limit for the nitrogen-to-iron
ratio is set by calculations which includseenhanced
convection prescriptions, but the lower limit is not defined
at low metallicities
since the standard TWW set of massive stars 
does not produce significant nitrogen yields. 
We see  from Figure~8 that there exists
a general consistency between the
observed [X/Fe] ratios and the theoretical predictions.
The P abundance is slightly suppressed whereas Ar behaves like other
$\alpha$-chain elements. 
However, the same consistency is not 
found if compared with the  halo stellar abundance pattern, 
where $\alpha$-chain elements 
are much more enhanced. There are  intrinsic uncertainties 
in the M~$\ge$ 30 M$_{\odot}$ extremely metal-poor massive-star models, which
may be important here. TWW suggest that their Fe yields may be overestimated by
a factor 2, which, if applied in Figure~8,  would shift all  abundances 
by 0.3 dex along the ordinates.
For instance in the Samland (1998) 
analysis the $\alpha$-chain elements are more
enhanced in agreement with what is observed in the Galactic halo stars.

The P abundance found in our DLA is slightly lower than Fe and the other iron-peak
elements  showing that P was slightly undersolar at low metallicities.
Since P is produced almost entirely by massive stars to achieve the higher 
solar value,  its production must increase with the metallicity of the gas
also to compensate the contribution to iron made by Type I SNae.
In other words the P yields have to be  necessarily metallicity dependent.
This  illustrates well how direct P observations in metal-poor environments
can be 
used to constrain the parameter values in the stellar
nucleosynthesis theory. 

The obtained results on P and Fe abundances do not seem to agree with the
dust formation processes in the Galaxy where iron depletion in the diffuse
interstellar clouds has been well studied. In Section 1, two possible
mechanisms of the dust formation processes were outlined, and it was
noted that P and Fe abundances may be used for choosing one of them. 
For instance,
if the dust grains grow in the interstellar medium we could detect
a disparity in  P and Fe gas-phase abundances as seen in the 
Galactic diffuse 
clouds.  
Obviously this is not the case for the \zabs = 3.3901 system. 
If, however, the dust grains are formed in hot shells of supernova 
remnants -- 
an
example of this process has recently been presented by Arendt et al. (1999) 
who 
revealed Fe protosilicate in the Cas~A ejecta --  then we should observe a 
gas-phase
depletion for both P and Fe, since they have the same condensation 
temperature.
This has not been found in our DLA either. Therefore, neither of the two
possible mechanisms of dust formation seem to work in the \zabs = 3.3901
damped Ly-$\alpha$ cloud. Taking into account that both volatile and
non-volatile elements do not show significant differences in their abundances
from the expected quantities, we may conclude that in this DLA possible 
destructive
processes are also insufficient  to cause some evaporation of the 
volatile substances.

\subsection{Neutral hydrogen  density}
 
The number density of neutral hydrogen, $n_{\rm HI}$, in
the \zabs = 3.3901 system can be estimated by 
using the relative population ratio of
the first excited state O\,{\sc i}$^\ast$ ($^3$P$_1$) 
to the ground state O\,{\sc i} ($^3$P$_2$). 

In H\,{\sc i} regions with  temperatures ranging 
from 100~K to 1000~K, the rate coefficients for
collisions with hydrogen atoms are larger than those for electron
collisions when the electron density, $n_{\rm e}$, is small, i.e.
$n_{\rm e} / n_{\rm HI} \la 10^{-2}$ (Bahcall \& Wolf 1968).
If we also neglect the contributions of 
proton and molecular hydrogen collisions, 
 the equilibrium equation for the three-level system
($^3$P$_2$, $^3$P$_1$, and $^3$P$_0$ states of O\,{\sc i})
can be written as
\begin{equation}
\frac{N({\rm O}\,{\sc i}^\ast)}{N({\rm O}\,{\sc i})} \simeq
\frac{n_{\rm H}\,(k_{01} + k_{02})}{A_{10}}\; ,
\label{eq:E1}
\end{equation}
where $k_{01}$ and $k_{02}$ are the rate coefficients of the
O--H collisions, and $A_{10} = 8.95\times10^{-5}$~s$^{-1}$ 
is the spontaneous transition probability
(all other collisional and photon rates are negligibly 
small in this case).

The rate coefficients for the downward transitions 
$k_{10}$ and $k_{20}$, induced by O--H
collisions in the temperature range 
from 50~K to 1000~K, are given by
Launay \& Roueff (1977). The corresponding
excitation rates for upward transitions can 
be calculated from the
principle of detailed balance which implies
\begin{equation}
k_{0i} = \frac{g_i}{g_0}\,\exp\left( 
- \frac{\Delta E_{0i}}{T}\right)\,k_{i0}\; ,
\label{eq:E2}
\end{equation}
where $g_2 = 1$, $g_1 = 3$ and $g_0 = 5$ 
are the degeneracies of states
$^3$P$_0$, $^3$P$_1$ and $^3$P$_2$, respectively.  
$\Delta E_{01} = 228$~K and $\Delta E_{02} = 326$~K are 
the energy differences between the ground state and the 
corresponding fine-structure levels.
 
In the observed spectrum of QSO~0000--2620 
the most stringent limit on the
population of the $^3$P$_1$ state may be set by
the O\,{\sc i}$^\ast\,\lambda990.2043$ \AA\, transition (see Figure~9).
Using the  $b$-parameter value 
of 10 km~s$^{-1}$, we estimate
$N({\rm O}\,{\sc i}^\ast) \leq 1.9\times10^{12}$ cm$^{-2}$, 
at 1 $\sigma$ confidence level.
Then, with the observed ratio
$N({\rm O}\,{\sc i}^\ast)/N({\rm O}\,{\sc i}) \leq 7.2\times10^{-5}$,
and assuming $T \simeq 10^3$~K (see Paper~II), we find
from Launay \& Roueff
$k_{10} = 9.22\times10^{-11}$ cm$^3$~s$^{-1}$ and
$k_{20} = 3.80\times10^{-11}$ cm$^3$~s$^{-1}$
or
$k_{01} = 4.4\times10^{-11}$ cm$^3$~s$^{-1}$ and
$k_{02} = 0.55\times10^{-11}$ cm$^3$~s$^{-1}$,
and get $n_{\rm HI} < 130$ cm$^{-3}$.
(Note that for the chosen temperature and for
$n_{\rm e}/n_{\rm HI} = 10^{-2}$ 
the rate coefficient for collisions with
electrons equals to $3\times10^{-12}$~cm$^3$~s$^{-1}$
if the collision strength is
$\Omega_{01} = 3.35\times10^{-3}$ according to Bell et al. 1998).

Since O\,{\sc i}$^\ast$ traces neutral phase regions along the line
of sight, the above result may be considered as the upper limit for
the whole H\,{\sc i} cloud. 
If this is the case, then the physical size of
the \zabs = 3.3901 cloud is $L > 7$~pc.
If the lower limit is $n_{\rm HI} > 1$~cm$^{-3}$, which is
typical for the Galactic diffuse H\,{\sc i} clouds, then
$L < 850$~pc. Comparable  sizes  have been also placed by 
Giardino \& Favata (2000) for the second damped
system at \zabs = 3.054 towards QSO~0000--2620 from the detection of the 
C\,{\sc ii} and C\,{\sc ii}$^\ast$ absorptions.

\section{Conclusions}
 
UVES observations of the quasar Q~0000--2620 gave as a result 
the first detection 
of the Ar\,{\sc i}\,$\lambda1066.660$ \AA\, and 
P\,{\sc ii}\,$\lambda963.801$ \AA\, transitions originated in the damped 
Ly$\alpha$ system at \zabs = 3.3901. 
The  derived abundances are 
  [Ar/H] = $-1.91 \pm 0.09$ and   [P/H] = $-2.31 \pm 0.10$
with most  errors originated in the neutral hydrogen column density.

The phosphorus abundance provided here is one of the few
available for this element in any astronomical site and unique
in  material poor of metals.
The low abundance of the non-refractory element P confirms 
the low  metallicity level of the  galaxy which is therefore in the early
stages of its chemical evolution.  
The P abundance is slightly below the iron-peak
elements and when
compared to its close nuclei S and Si, P shows mild 
evidence of an odd-even effect, but less extreme than what predicted by 
theoretical models
by Woosley \& Weaver (1995) or Limongi et al. (2000).

Ar abundance is remarkably similar to that of 
oxygen, sulphur and silicon, 
[O/H] = $- 1.86 \pm 0.15$, 
[S/H] = $- 1.98 \pm 0.09$, and 
[Si/H] = $- 1.91 \pm 0.08$.
Ar has a large photoionization cross-section and the similarity
of its abundance with the other $\alpha$-chain elements 
implies the absence of 
significant photoionization  either by external or internal sources.
 
Log~(Ar/O) and $\log$(Ar/S) ratios are almost  
the same as  those found
in the blue compact galaxies and in present day galaxies
where  O  is more abundant by one or two orders of magnitude. 
This strengthens the constancy of  Ar/O and Ar/S 
ratios and lends support to the existence of a universal IMF.

In the DLA under study 
the relative abundances of $\alpha$-chain and iron-peak 
elements, as measured by an unprecedented  number of indicators, 
show only a mild overabundance at variance with 
the pattern of the old stellar population of the Galaxy.
Since this DLA is a dust-free system, its chemical
composition and the relative abundances of the observed
species may be used to constrain model parameters in
the theoretical calculations of stellar nucleosynthesis
in the metal-poor gas.
 
Measurements of the volatile and nonvolatile element
abundances have shown that none of the known processes of
dust formation or destruction is effective  in the 
\zabs = 3.3901 galaxy.

Finally, the upper limit to the population of the
first fine-structure level in the ground state of 
O\,{\sc i} 
is used to infer a lower limit to 
the physical dimension of the 
DLA system, which is placed at $L > 7$~pc.

\section{Acknowledgments}
The  high resolution spectra analyzed in this paper 
are of unique
quality and were obtained during the first nights of 
commissioning of a new instrument at a new telescope. 
For these results
we are indebted to all ESO staff involved in the 
VLT construction and in 
UVES commissioning. We thank also the referee Jason X. Prochaska for his 
useful comments and remarks.
The work of S.A.L. is supported by the RFBR grant
No. 00--02--16007.

\clearpage

\psfig{figure=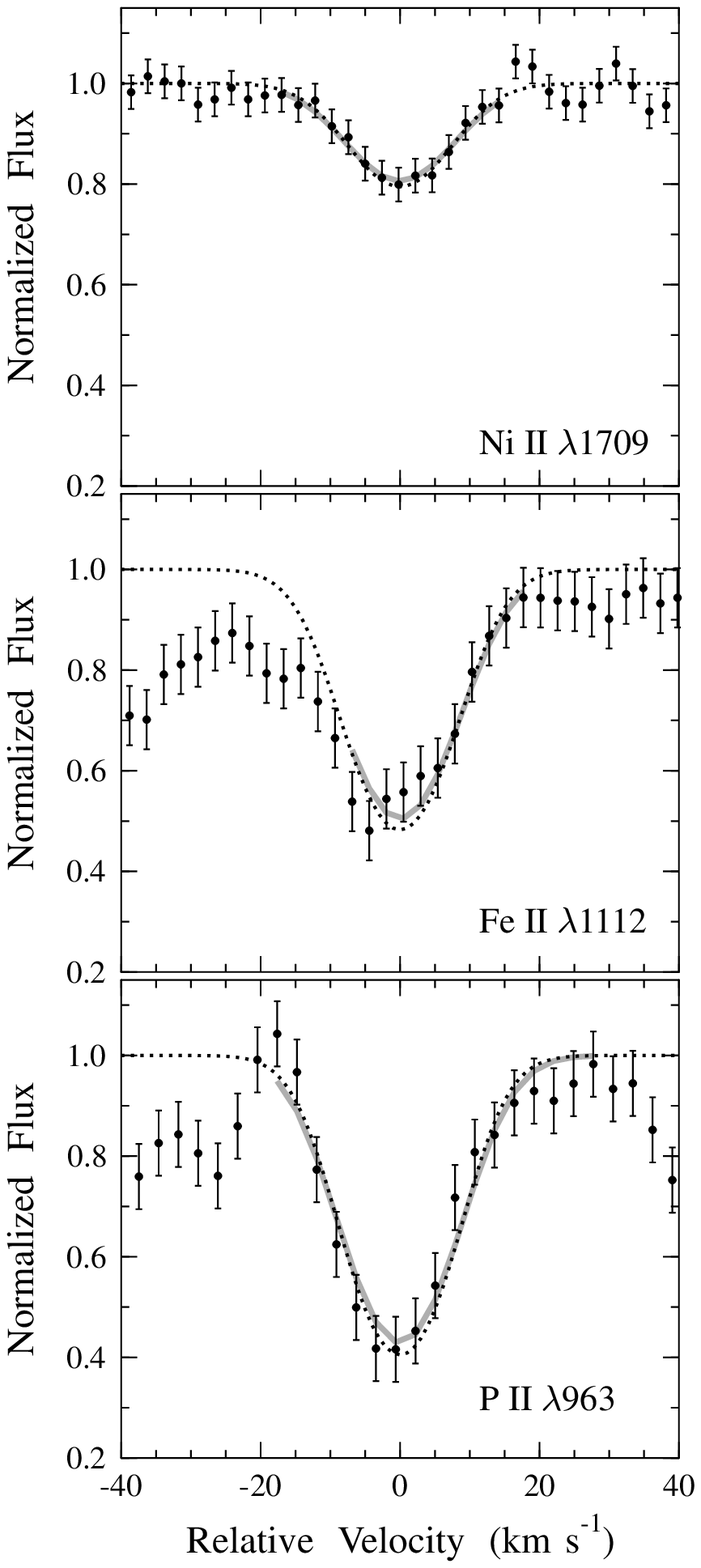,clip=t}
\figcaption[f1.ps]{Ni\,{\sc ii}\,$\lambda1709.600$ \AA\ (top panel),
Fe\,{\sc ii}\,$\lambda1112.048$ \AA\ (middle panel) and
P\,{\sc ii}\,$\lambda963.801$ \AA\ (bottom panel) lines associated with the 
\zabs = 3.3901 DLA system towards QSO~0000--2620 
(dots and $1\sigma$ error bars). 
The lines  are aligned taking zero radial velocity
in correspondence to \zabs = 3.390127.
Smooth lines are the synthetic spectra obtained from the fit (see text). 
}
\label{fig1} 
\vskip 5 truecm 
\psfig{figure=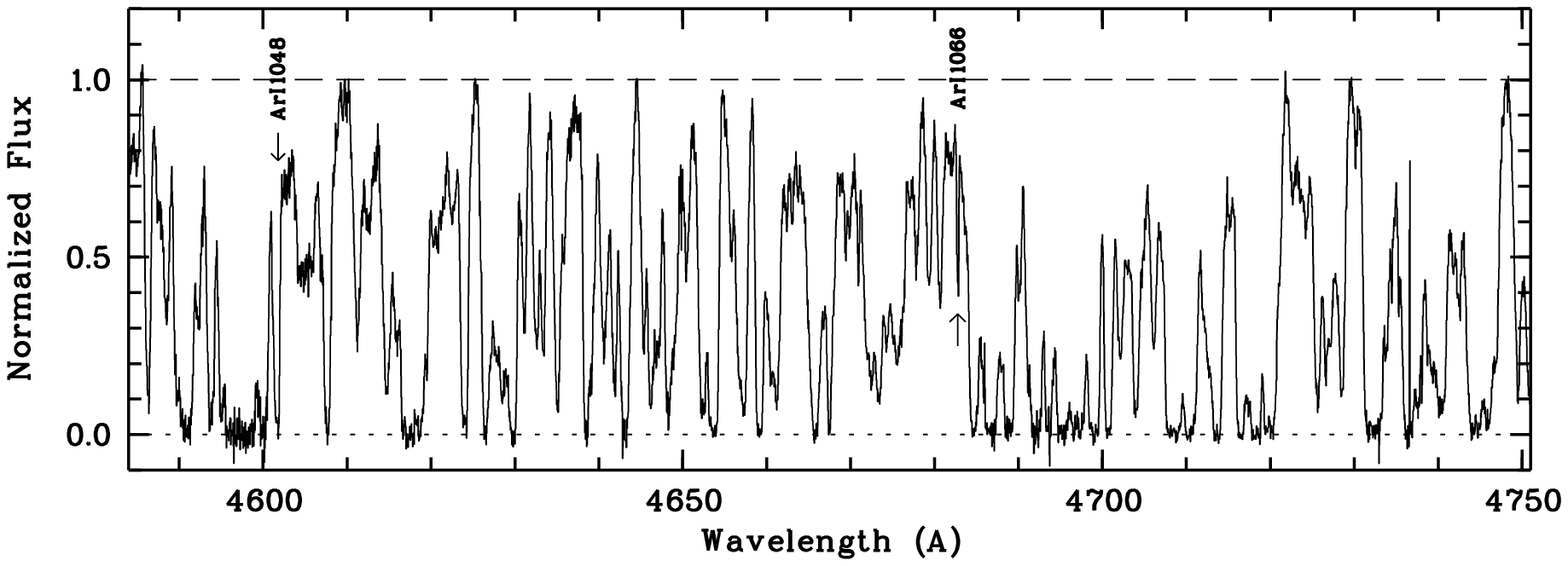,angle=-90,clip=t}
\figcaption[f2.ps]{
A portion of the normalized spectrum of QSO 0000--2620
in the region of the Ar\,{\sc i}\,$\lambda1048.220$ \AA\ and 
Ar\,{\sc i}\,$\lambda1066.660$ \AA\ lines 
($\lambda_{\rm obs} = 4601.8$ \AA\ and 4682.8 \AA, respectively)
associated with the \zabs = 3.3901 DLA.
All the highest windows in the forest have been used for
continuum tracing as well as other windows outside the portion
displayed in a spectral range spanning about 1000 \AA.
}
\label{fig2}  
\psfig{figure=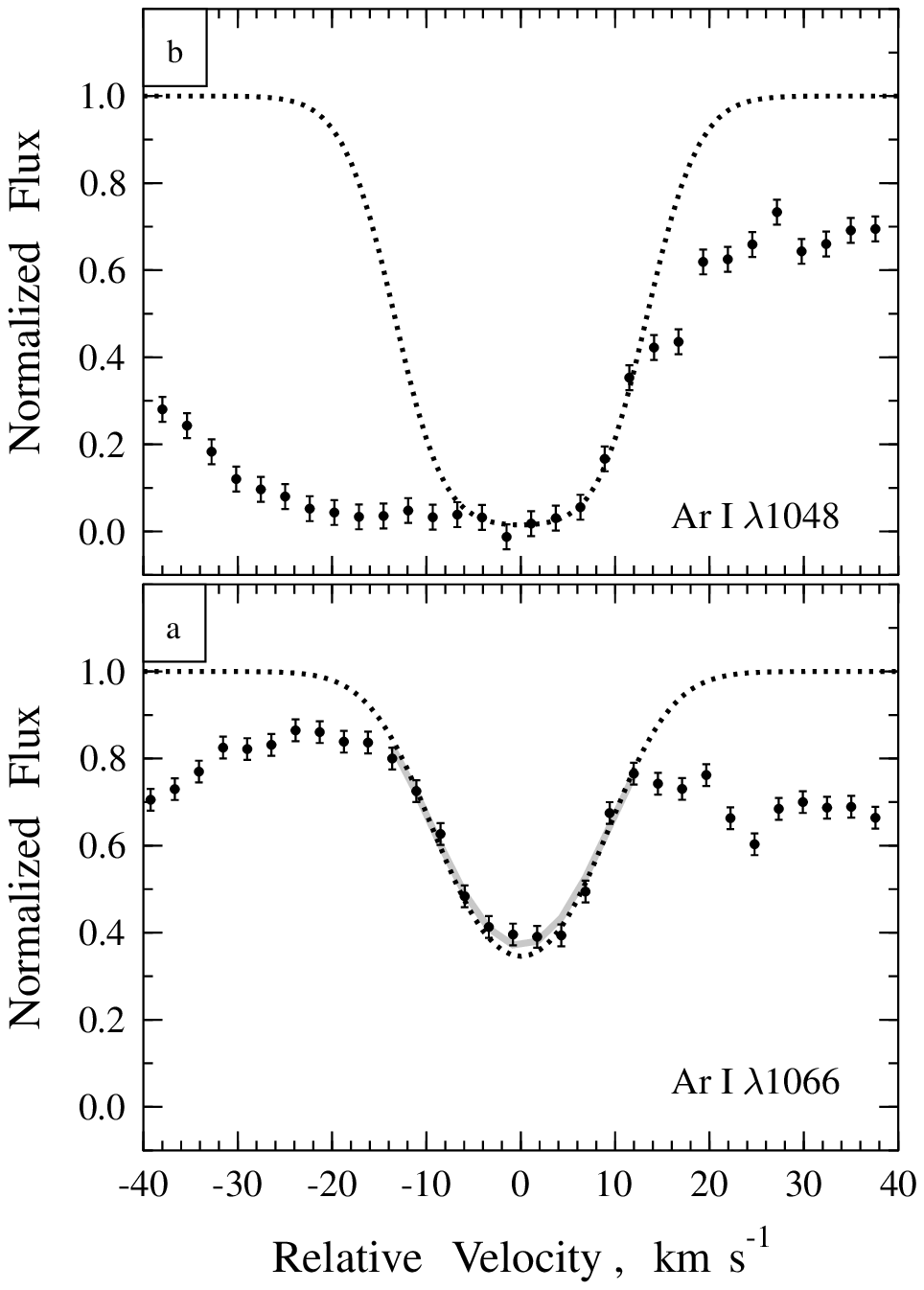,clip=t}
\figcaption[f3.ps]{ ( {\bf a})
Ar\,{\sc i}\,$\lambda1066.660$ \AA\, line associated with the 
\zabs = 3.3901 DLA system towards QSO~0000--2620 (dots and error bars). 
The line is aligned taking zero radial velocity
in correspondence to \zabs = 3.390127.
Smooth lines are the synthetic spectra obtained from the fit  
as described in the text. 
({\bf b}) A consistent fit of the red wing of the
Ar\,{\sc i}\,$\lambda1048.220$ \AA\ line (the dotted curve)
based on the best fit of the Ar\,{\sc i}\,$\lambda1066$ \AA\
line shown in panel   {\bf a} by the dotted curve as well.
}
\label{fig3}  
\psfig{figure=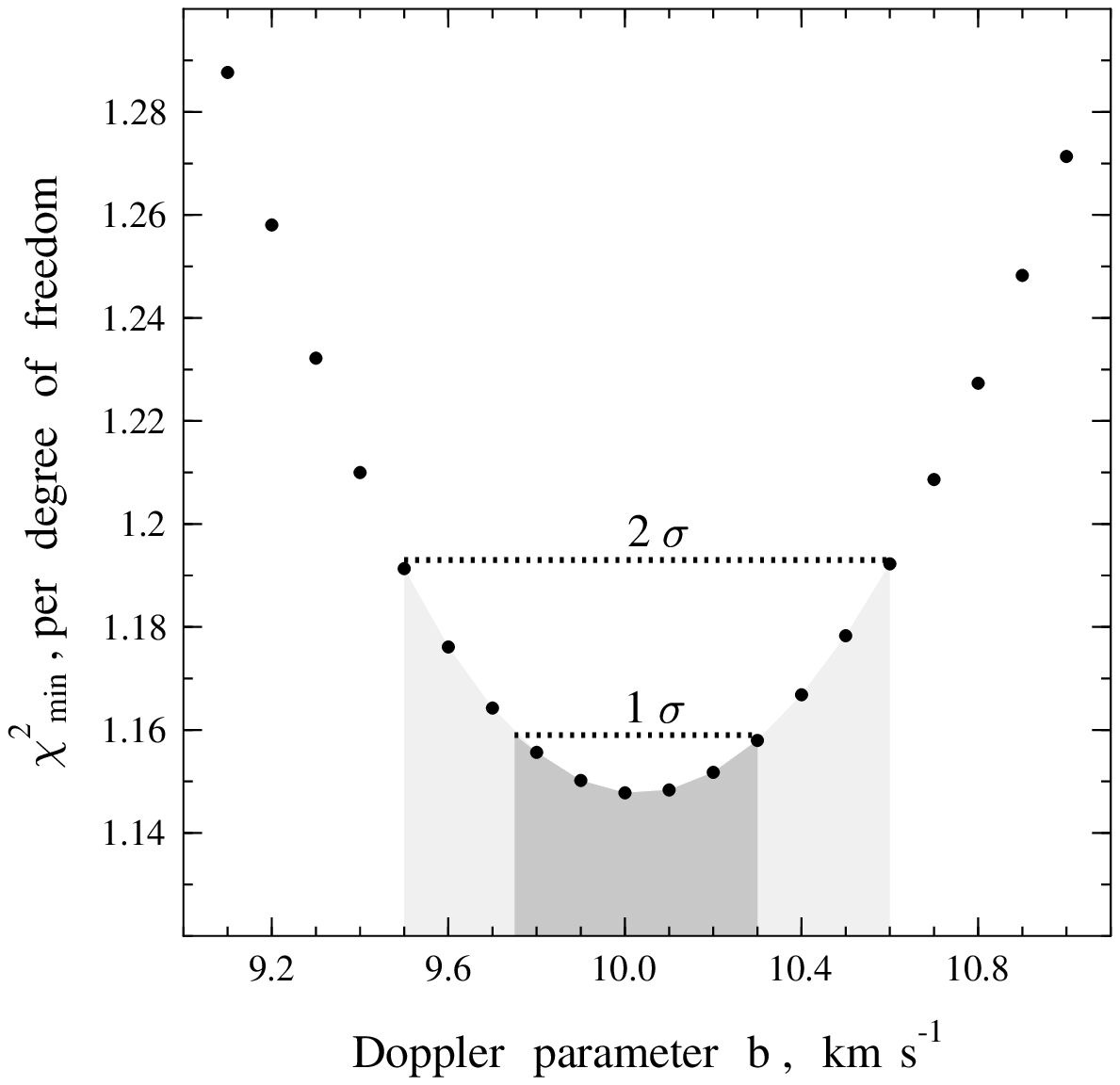,clip=t}

\figcaption[f4.ps]{ 
Confidence regions in the  $\chi^2_{\rm min}$ -- $b$ plane
calculated from the simultaneous fit of the metal absorption
lines located redwards the Ly$\alpha$ emission~:
Fe\,{\sc ii}\,$\lambda1611$, Ni\,{\sc ii}\,$\lambda1709$,
Si\,{\sc ii}\,$\lambda1808$, Zn\,{\sc ii}\,$\lambda2026$,
Cr\,{\sc ii}\,$\lambda2056$, and Cr\,{\sc ii}\,$\lambda2062$.
The parabola vertex corresponds to the best value of
$b = 10$ \kms.
}
\label{fig4}  
\psfig{figure=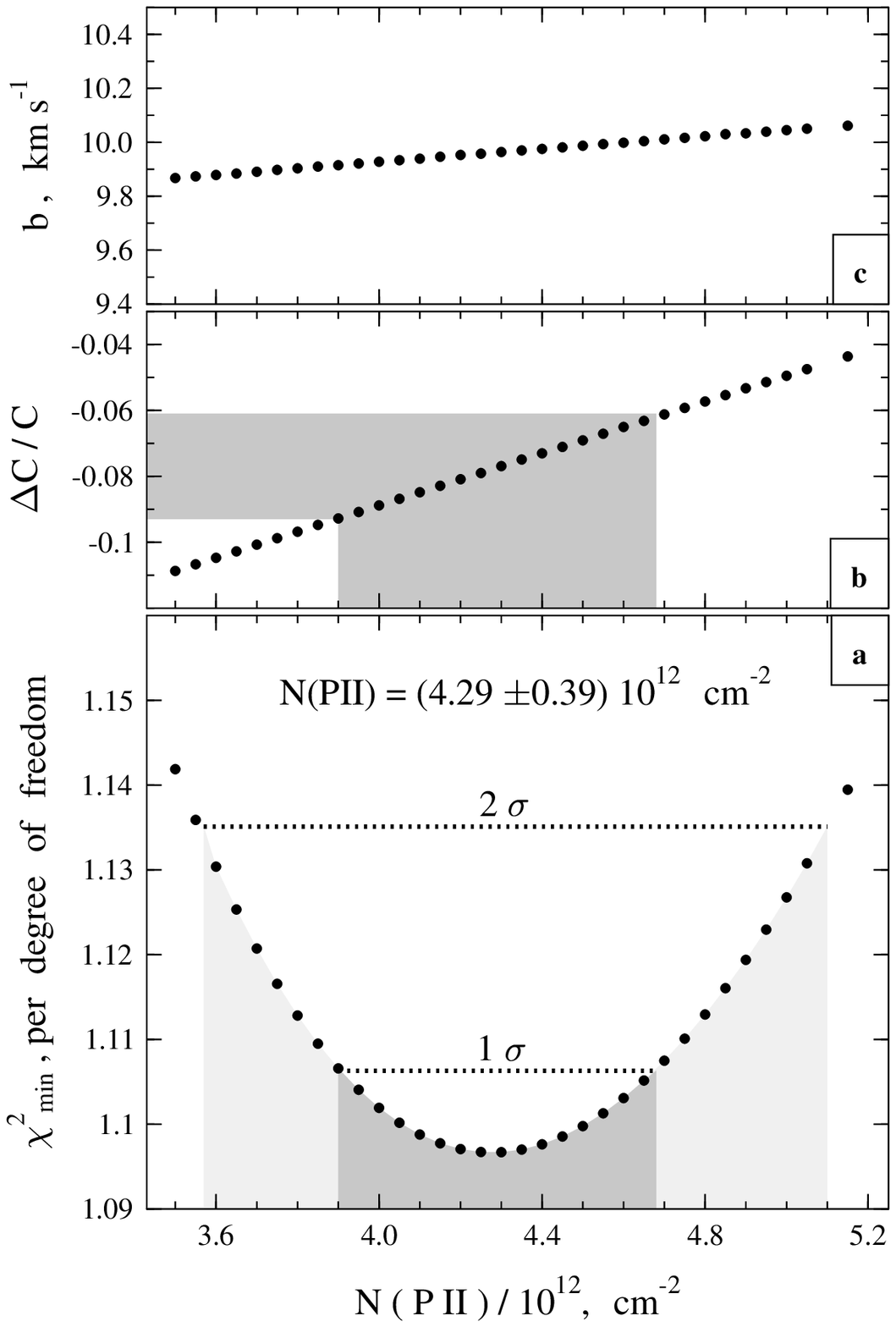,clip=t}
\figcaption[f5.ps]{ 
Results of the fitting procedure for the P\,{\sc ii}\,$\lambda963$ \AA\
line. ({\bf  a}) The $\chi^2_{\rm min}$ as a function of the column
density of the singly ionized phosphorus calculated from
the simultaneous fit of P\,{\sc ii}\,$\lambda963$ 
and metals
Fe\,{\sc ii}\,$\lambda1611$, Ni\,{\sc ii}\,$\lambda1709$,
Si\,{\sc ii}\,$\lambda1808$, Zn\,{\sc ii}\,$\lambda2026$,
Cr\,{\sc ii}\,$\lambda2056$, Cr\,{\sc ii}\,$\lambda2062$.
The 68.3\%  and 95.4\%  confidence levels are marked
(the lower and upper horizontal dotted lines, respectively).
The parabola vertex corresponds to the best value of
$N({\rm P}\,{\sc ii}) = 4.29\times10^{12}$ cm$^{-2}$.
(  b) The $\Delta C/C$ variations of the local continuum level.
The grey area restrics the $1\sigma$ deviations of $\Delta C/C$
in accordance to panel   {\bf a}).  ({\bf   c}) The corresponding values
of the Doppler $b$-parameter.
}
\label{fig5}  
\clearpage
\psfig{figure=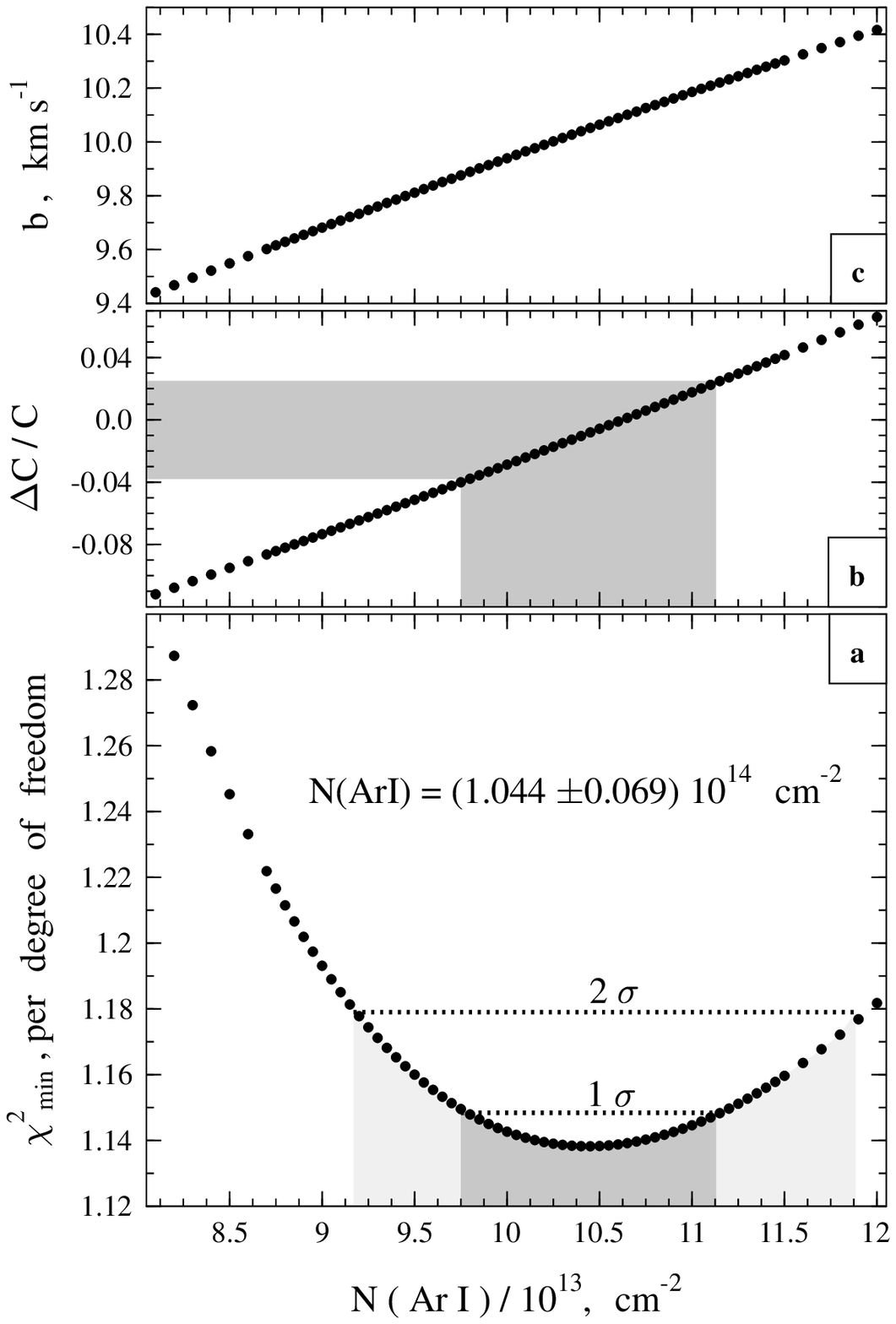,clip=t}
\figcaption[f6.ps]{
As Fig.~5 but for the Ar\,{\sc i}\,$\lambda1066$ \AA\ line.
The parabola vertex in panel {\bf a}
corresponds to the best value of 
$N({\rm Ar}\,{\sc i}) = 1.044\times10^{14}$ cm$^{-2}$.
}
\label{fig6}  

\psfig{figure=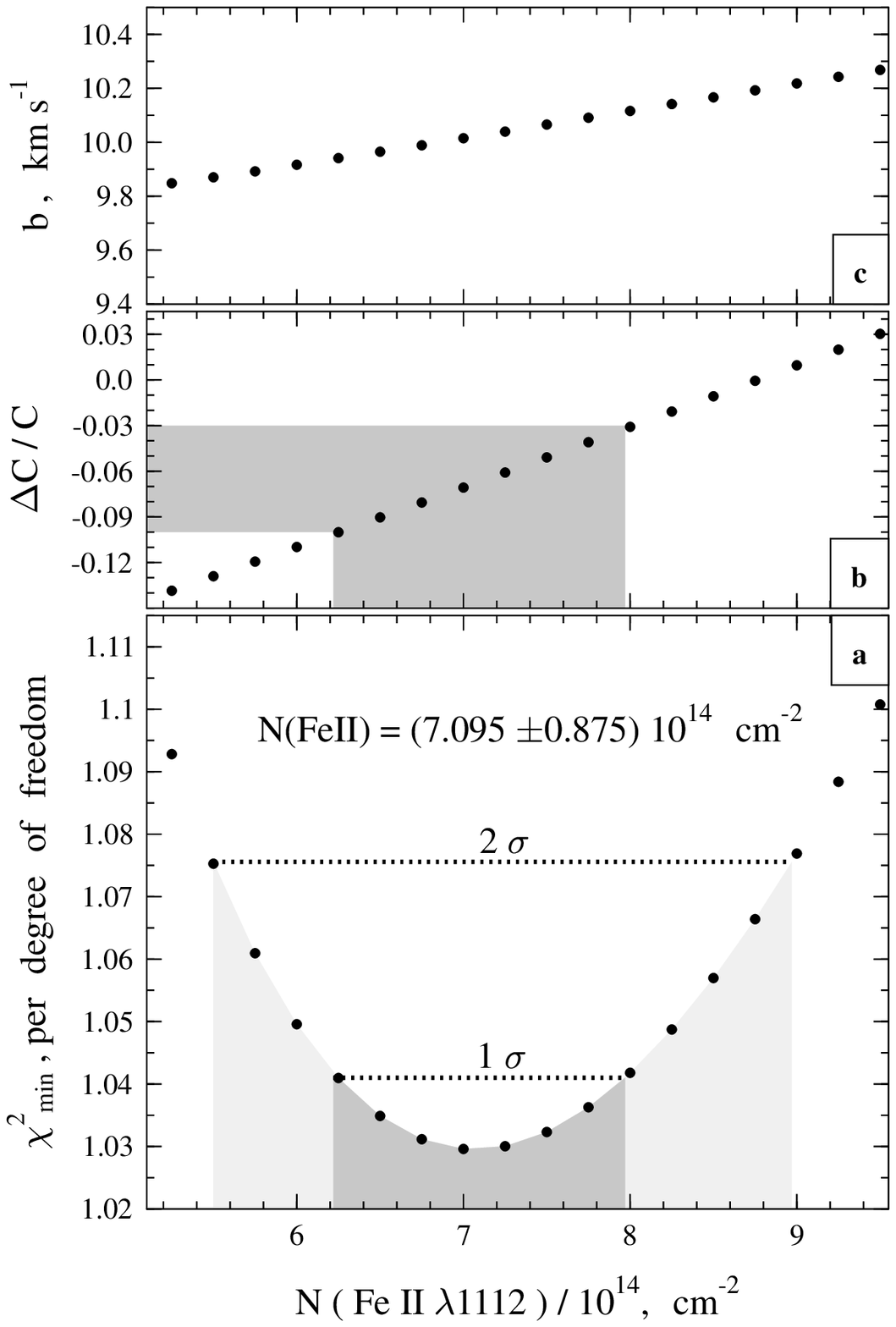,clip=t}
\figcaption[f7.ps]{
As Fig.~5 but for Fe\,{\sc ii}\,$\lambda1112$ \AA\ line
and the following set of the metal absorption lines
from the red portion of the quasar spectrum~:
Ni\,{\sc ii}\,$\lambda1709$,
Si\,{\sc ii}\,$\lambda1808$, Zn\,{\sc ii}\,$\lambda2026$,
Cr\,{\sc ii}\,$\lambda2056$, Cr\,{\sc ii}\,$\lambda2062$.
The parabola vertex in panel {\bf a}
corresponds to the best value of 
$N({\rm Fe}\,{\sc ii}\,\lambda1112) = 7.095\times10^{14}$ cm$^{-2}$.
}
\label{fig7}  
\psfig{figure=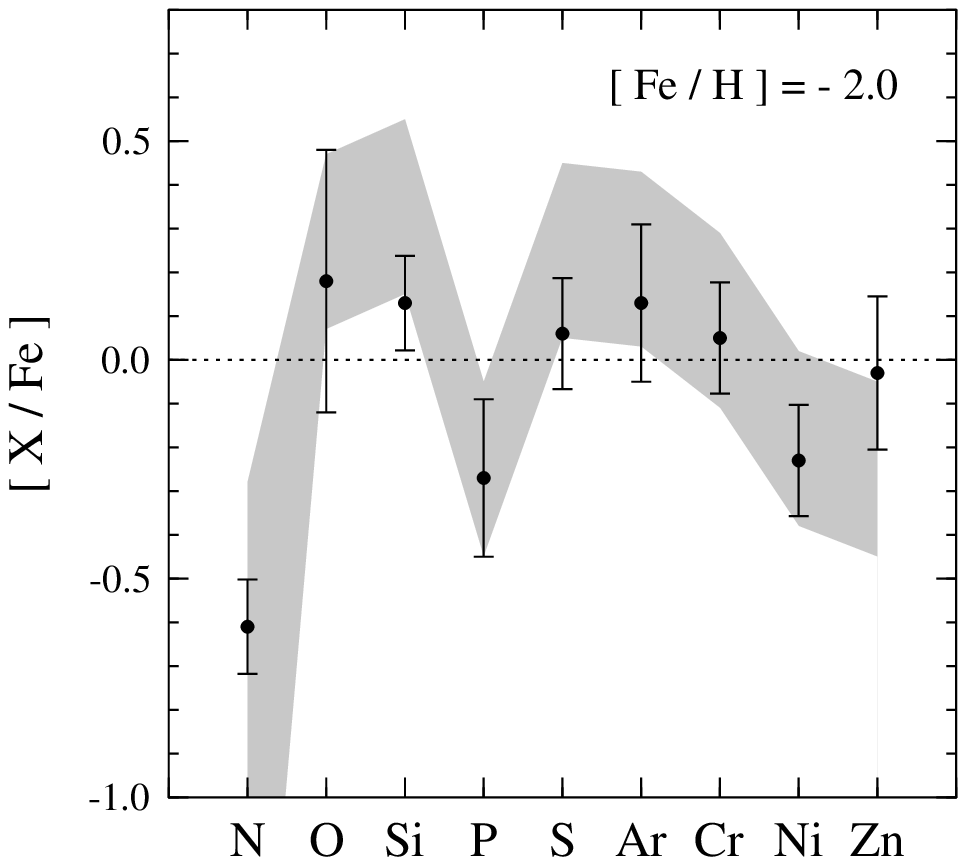,clip=t}
\figcaption[f8.ps]{
The theoretical nucleosynthetic yields (the shaded area
shows $\pm 0.2$ dex uncertainty range, see text for more
details) for the metallicity [Fe/H] = -- 2.0 in the
standard model of the galactic chemical evolution by
Timmes et al. (1995). The dots with 3$\sigma$ error bars mark
the measured relative abundances [X/Fe] at \zabs = 3.3901
towards QSO~0000--2620. }
\label{fig8} 
\clearpage 
\psfig{figure=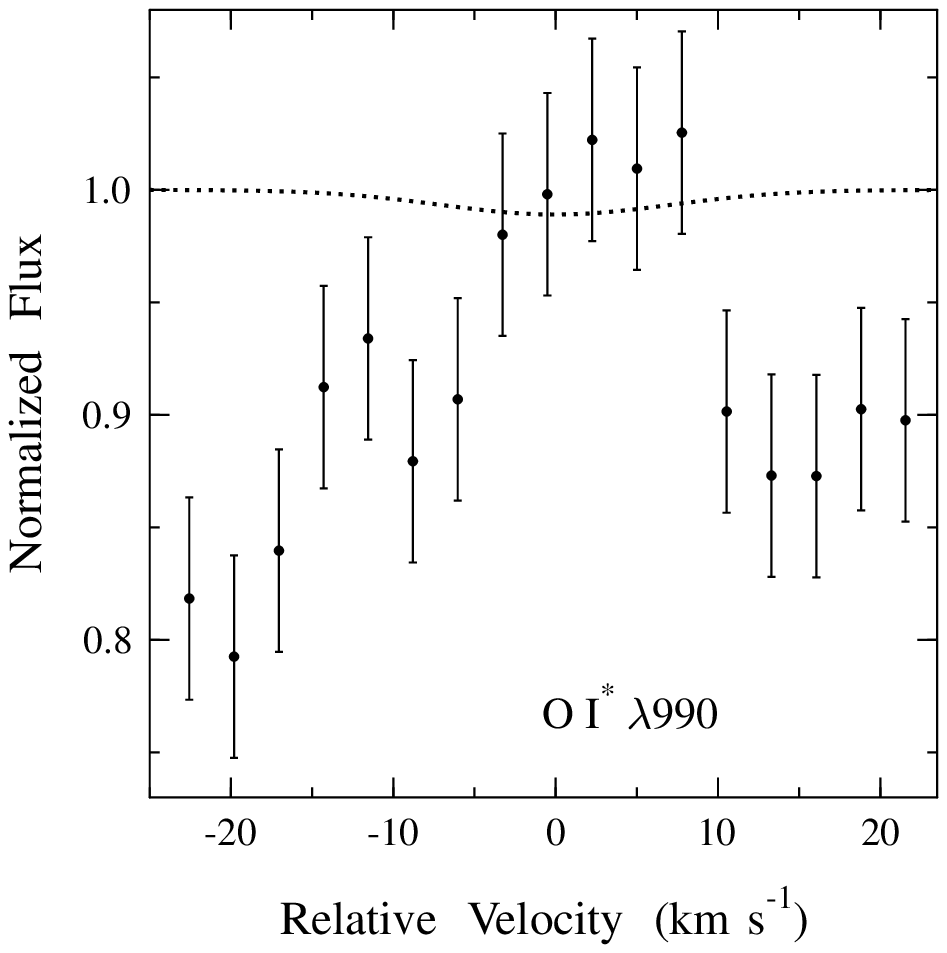,clip=t}
\figcaption[f9.ps]{
O\,{\sc i}$^{\ast}\,\lambda990.2043$ \AA\ 
(transition $^3{\rm P}_1 \to$ $^3{\rm D}_2$)
expected position in the 
\zabs = 3.3901 DLA system towards QSO~0000--2620.  
The radial velocity scale is the same as in Figs.~1 and 3.
The dotted curve is the 
convolved with the spectrograph function
synthetic spectrum constraining
the total O\,{\sc i}$^\ast$ column density for the
fixed value of $b = 10$ \kms.
}
\label{fig9}

\clearpage

\pagestyle{empty}
\begin{deluxetable}{ccccc}
\footnotesize
\tablewidth{-5cm} 
\tablecaption{Column densities and abundances in the dust-free
pattern at \zabs = 3.3901} 
\tablehead{
\colhead{Element} & \colhead{$\log\,(N)$\,$\tablenotemark{a}$} &
\colhead{(X/H)$_{\rm DLA}$} & 
\colhead{(X/H)$_\odot$$\tablenotemark{b}$} & 
\colhead{[X/H]\,$\tablenotemark{c}$}  
} 
\startdata
$^1_1$H & $21.41 \pm 0.08\,\tablenotemark{d}$ &  &  & \nl     
$^{14}_{\,\,\,7}$N & $14.73 \pm 0.02\,\tablenotemark{g}$ &
$-6.68 \pm 0.083$ & $-4.03 \pm 0.07$ & 
$-2.65 \pm 0.12\,\tablenotemark{e}$ \nl
$^{16}_{\,\,\,8}$O & $16.42 \pm 0.10\,\tablenotemark{g}$ &
$-4.99 \pm 0.128$ & $-3.13 \pm 0.07$ &  
$-1.86 \pm 0.15\,\tablenotemark{e}$ \nl
$^{28}_{14}$Si & $15.06 \pm 0.02$ &
$-6.35 \pm 0.083$ & $-4.44 \pm 0.01$ & 
$-1.91 \pm 0.08\,\tablenotemark{e}$ \nl
$^{31}_{15}$P & $12.63 \pm 0.04$ & 
$-8.78 \pm 0.089$ & $-6.47 \pm 0.04$ & 
$-2.31 \pm 0.10\,\tablenotemark{f}$ \nl
$^{32}_{16}$S & $14.70 \pm 0.03$ &
$-6.71 \pm 0.085$ & $-4.73 \pm 0.01$ & 
$-1.98 \pm 0.09\,\tablenotemark{d}$ \nl
$^{36}_{18}$Ar & $14.02 \pm 0.03$ &
$-7.39\pm 0.085$ & $-5.48 \pm 0.04$ & 
$-1.91 \pm 0.09\,\tablenotemark{f}$ \nl
$^{52}_{24}$Cr & $13.10 \pm 0.03\,\tablenotemark{g}$ &
$-8.31 \pm 0.085$ & $-6.32 \pm 0.03$ & 
$-1.99 \pm 0.09\,\tablenotemark{e}$ \nl
$^{56}_{26}$Fe & $14.87 \pm 0.03$ &
$-6.54 \pm 0.085$ & $-4.50 \pm 0.01$ & 
$-2.04 \pm 0.09\,\tablenotemark{e}$ \nl
$^{58}_{28}$Ni & $13.39 \pm 0.03$ &
$-8.02 \pm 0.085$ & $-5.75 \pm 0.01$ & 
$-2.27 \pm 0.09\,\tablenotemark{e}$ \nl
$^{64}_{30}$Zn & $12.01 \pm 0.05$ &
$-9.40 \pm 0.094$ & $-7.33 \pm 0.04$ & 
$-2.07 \pm 0.10\,\tablenotemark{e}$ \nl
\enddata
\tablenotetext{}{NOTES -- 
(a) $N$ in cm$^{-2}$, (b) data from Grevesse et al. (1996)
except for Ar where the weighted average values from
Sofia \& Jenkins (1998) are used,
(c) errors in [X/H] include errors in column densities 
and in solar abundances,
(d) from Lu et al. (1996), (e) from Molaro et al. (2000),
(f) this work, (g) mean values.  
}
\end{deluxetable}

\clearpage 

\pagestyle{empty}
\begin{deluxetable}{ccccc}
\footnotesize
\tablewidth{-5cm} 
\tablecaption{The [O,Si,S,Ar/Cr,Fe,Ni,Zn] ratios in the
\zabs = 3.3901 system}
\tablehead{
\colhead{ } & \colhead{Cr} &
\colhead{Fe} & \colhead{Ni} & \colhead{Zn}  
} 
\startdata
O&$0.13\pm0.10$&$0.18\pm0.10$&$0.41\pm0.10$&$0.21\pm0.11$ \nl
Si&$0.08\pm0.03$&$0.13\pm0.04$&$0.36\pm0.04$&$0.16\pm0.05$ \nl
S&$0.01\pm0.04$&$0.06\pm0.04$&$0.29\pm0.04$&$0.09\pm0.06$ \nl
Ar&$0.08\pm0.04$&$0.13\pm0.06$&$0.36\pm0.05$&$0.16\pm0.06$ \nl
\enddata
\tablenotetext{}{}
\end{deluxetable}

\end{document}